# Control water waves by metagratings

Linkang Han, Qilin Duan, Junliang Duan, Shan Zhu, Shiming Chen, Yuhang Yin and Huanyang Chen*

Department of Physics, Xiamen University, Xiamen, 361005, China

*Abstract.* Metasurfaces and metagratings offers new platforms for electromagnetic wave control with significant responses. However, metasurfaces based on abrupt phase change and resonant structures suffer from the drawback of high loss and face challenges when applied in water waves. Therefore, the application of metasurfaces in water wave control is not ideal due to the limitations associated with high loss and other challenges. We have discovered that non-resonant metagratings exhibit promising effects in water wave control. Leveraging the similarity between bridges and metagratings, we have successfully developed a water wave metagrating model inspired by the Luoyang Bridge in ancient China. We conducted theoretical calculations and simulations on the metagrating and derived the equivalent anisotropic model of the metagrating. This model provides evidence that the metagrating has the capability to control water waves and achieve unidirectional surface water wave. The accuracy of our theory is strongly supported by the clear observation of the unidirectional propagation phenomenon during simulation and experiments conducted using a reduced version of the metagrating. It is the first time that the unidirectional propagation of water waves has been seen in water wave metagrating experiment. Above all, we realize the water wave metagrating experiment for the first time. By combining complex gratings with real bridges, we explore the physics embedded in the ancient building—Luoyang Bridge, which are of great significance for the water wave metagrating design, as well as the development and preservation of ancient bridges.

*Introduction.* Water waves are ubiquitous in our daily lives, representing one of the most common forms of waves. They are mechanical waves that widely exists on the surface of various waters, and its restoring force is provided by gravity [1,2]. The investigation of water wave scattering, diffraction, and decay as well as the development of novel techniques for water waves control are hot topics in the field of water wave mechanics for a long time [2,3]. In the past ten years, photonic crystals [4-7], metamaterials [8-11] and metasurfaces/metagratings [12-16] have demonstrated significant success in the control of electromagnetic waves. Consequently, researchers have extended these structures and methods into water wave systems (new corresponding relationship can be seen Supplemental Material, Sec. 1[17]), such as employing photonic crystals to achieve negative refraction [18] and self-collimation [19] of water waves, etc. Additionally, the utilization of metamaterials has enabled advancements in water wave control, including cloaking [20,21], rotation of water waves [22], focusing [23,24] and concentration [25], etc. Furthermore, periodic metamaterials have shown promise in achieving extraordinary effects in water wave systems, which can be applied to achieve negative gravity [44] and surpass the upper limit of water wave velocity [26], et al. However, the application of metasurfaces and metagratings in water waves control, especially in experiments, is still limited and infrequent. In our previous work, we have theoretically proved that the metagratings can effectively isolate water waves and provide protective measures for buildings [27], while no experiments have been conducted so far. Therefore, conducting experiments on water wave metagratings is a highly worthwhile and timely compelling topic for exploration.

Metasurfaces and metagratings are a new type of artificial subwavelength materials that have been very popular in recent years, which provide a new method for wave field control and enabling the implementation of interesting functions [12-16]. Some notable functions include extraordinary optical diffraction [28], asymmetric Perfect Diffraction [29], negative diffraction [30], large numerical aperture focusing [31], broadband holographic display [32], holographic photoelectric detection [33], angular asymmetric absorption [34], non-scattering manipulation of abnormal reflection and refraction [35-37], reversal of transmission and reflection [38], extend the incident angle to several discrete angles [39] and so on. Most metasurfaces are typically designed based on abrupt phase change and resonance, however, the propagation of water waves is often associated with substantial loss. Thus, the utilization of abrupt phase changes and resonant structures in metasurfaces will further deepen the loss of the water wave, resulting in poor water wave control effect. This may be a crucial reason why metasurfaces have not been extensively employed for water wave control. However, through our research, we have discovered that certain non-resonant metagratings exhibit minimal loss for water waves while demonstrating good control capabilities. In this letter, building upon the previous theoretical work, specifically the utilization of the Luoyang Bridge metagrating model for water wave control [27], we proceed to design a novel non-resonant metagrating model to investigate and comprehend the underlying mechanism of controlling water waves through metagratings.

Luoyang Bridge is an ancient bridge located in Quanzhou, China (see Fig. 1a [40]), which has a history of more than 1000 years, it stretches over 1000 meters and features an array of 46 pillars [41]. Remarkably, the unique ship-like pillar structures give Luoyang Bridge some unique physical characteristics, such as eliminating water waves [27,42]. Through the research of the metagrating model of Luoyang Bridge, we find another intriguing phenomenon: vortex water waves can excite unidirectional surface water wave near the metagrating, as shown in Fig. 1b.

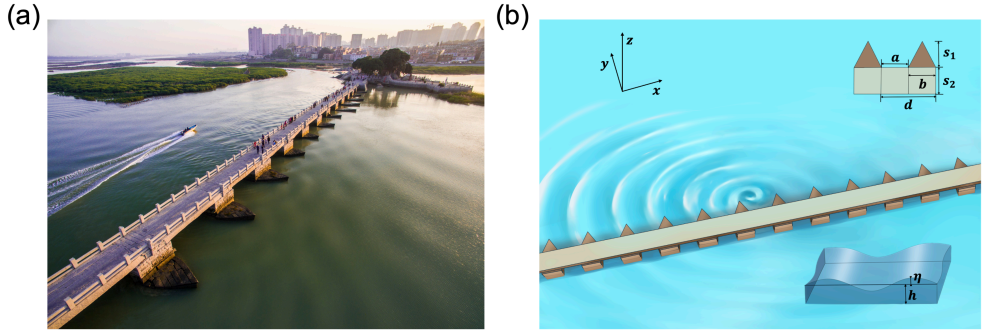

**FIG. 1.** Luoyang Bridge and its metagrating model. (a) Realistic terrain scene of Luoyang Bridge [40]. (b) The vortex water source excites a unidirectional surface water wave near the metagrating (the detailed structure diagram of the metagrating is shown at the upper right corner of the picture and the enlarged view of the vertical displacement of water surface $\eta$ and the static water depth $h$ (or average water depth) is shown at the lower right corner).

We can see that the vertical displacement of the vortex water wave (excited by the rotating propeller) at the right side of the source is greatly weakened, while at the left side of the source is enhanced, forming a unidirectional surface water wave. To comprehend the origins of this phenomenon, we conduct rigorous theoretical calculations and numerical simulations. Furthermore, we validate this phenomenon in experiments.

***Theoretical analysis and numerical simulation.*** If we consider linear, inviscid, and irrotational water waves in an infinite extent of water of constant depth $h$, its governing equation is [2,43]:

$$\nabla \cdot (u \cdot \nabla p) + \frac{\omega^2}{g} p = 0 \qquad (1)$$

where $u$ is reduced water depth, $u = tank(kh)/k$ [44], $p$ is the hydrostatic pressure of water surface ($p = \rho g \eta$), $\rho$ is the fluid density, $g$ is the gravitational acceleration, $\eta$ is the vertical displacement of the water wave, the nonlinear dispersion of water waves is [45]:

$$\omega = \sqrt{guk} \qquad (2)$$

where $\omega$ is the angular frequency, $k$ is propagation wave number of water wave.

Then we first write the vertical displacement expression of the vortex water wave in the $x - y$ plane as $\eta(x,y) = H_1(k_0 r)e^{i\theta} = \int \tilde{\eta}(k_x, y) e^{ik_x x} dk_x$, where

$$\tilde{\eta}(k_x, y) = \frac{1}{\pi k_0}\left[i\frac{k_x}{k_y} \mp 1\right] e^{ik_y |y - y_{source}|} \qquad (3)$$

$\tilde{\eta}(k_x, y)$ is spatial Fourier transform of $\eta(x,y)$, $H_1$ is the Hankel function of the first kind, $k_x$ $k_y$ and are the wave number in the $x$ and $y$ direction, $k_y = (k_0^2 - k_x^2)^{1/2}$ and $k_0 = 2\pi/\lambda$. Here $\lambda$ is the wavelength, $r$ and $\theta$ are cylindrical coordinate systems, and $r\cos\theta = x$, $r\sin\theta = y$. The minus and plus signs in the equation (3) respectively correspond to $y > y_{source}$ and $y < y_{source}$, where $y_{source}$ is the location of the source (e.g., a vortex water wave).

We assume that the wave number of the water wave propagating along the positive direction and negative direction of $x$-axis are $k_x > 0$ and $k_x < 0$ respectively. Because the wave source is located at $(0, y_{source})$ in the Fig. 1b, we can use the left part and right part of the wave source to represent $k_x < 0$ and $k_x > 0$ respectively.

When $|k_x| > k_0$, $k_y = i\sqrt{k_x^2 - k_0^2}$, then we bring it into (Eq.3) to get:

$$\tilde{\eta}(k_x, y) = \frac{1}{\pi k_0}\left[\frac{k_x}{\sqrt{k_x^2 - k_0^2}} \mp 1\right] e^{-\sqrt{k_x^2 - k_0^2}|y - y_{source}|} \qquad (4)$$

It can be seen from equation (4) that $\eta$ is an evanescent wave in the $y$ direction, and in the region of $y < y_{source}$ (corresponding to plus signs in the equation (4)), the vertical displacement of the evanescent wave components are enhanced in the part of $k_x > 0$ (on the right part of wave source), while for the part of $k_x < 0$ (on the left part of wave source), the evanescent wave components are mutually eliminated. Therefore, by making the propagation wave number of the evanescent wave in the $x$ direction meeting $|k_x| > k_0$, it is possible to excite the unidirectional surface water waves [46]. Does the metagrating can excite surface wave with $|k_x| > k_0$? Let's analyze and calculate it below.

Due to the complex structure of metagrating shown in Fig. 1b, it is difficult to carry out accurate analytical calculation at present. We first simplify it into a structure of periodic rectangular grating (which is directly called SPRG later) as shown in Fig. 2a, and the top view of the specific dimensions of the two adjacent rectangles is shown in the upper right corner of Fig. 2a, the length of each rectangle is $s_1 + s_2 = 11\,m$, width is $b = 5\,m$, and the distance between two rectangles is $a = 6\,m$, that is, the length of a unit cell is $d = a + b = 11\,m$. We first solve the dispersion relation of the excited surface mode of SPRG as: [47] (the derivation is in the Supplementary Materials

Section One):

$$k_x = \frac{\sqrt{d^2(e^{ik_0(s_1+s_2)} + 1)^2 - a^2(e^{ik_0(s_1+s_2)} - 1)^2}}{d(1 + e^{ik_0(s_1+s_2)})} k_0 \tag{5}$$

Through calculations, we find that SPRG can excite the surface mode of water waves, localize the water waves on its surface, form surface water waves, so the $x$-direction propagation wave number $|k_x|$ will become large, exceeding $k_0$, thereby unidirectional propagation of water waves can be realized. Analytical calculation and numerical simulation verification are carried out below.

We cannot simulate the infinitely long SPRG in a limited space, so we can make the SPRG be equivalent to an anisotropic water layer, if the reduced water depth in the area of $y > -6m$ and $y < -17m$ in Fig. 2c is set as $u_0$, then the equivalent parameters of the anisotropic water layer is $u_x = 0, u_y = a/d * u_0, g = d/a * g_0$ (the derivation is in the Supplementary Materials Section One). Then, the dispersion relation of the excited surface mode of the anisotropic water layer can be solved as follow (the derivation is in the Supplementary Materials Section One):

$$k_x = \frac{\sqrt{d^2(e^{ik_0(s_1+s_2)} + 1)^2 - a^2(e^{ik_0(s_1+s_2)} - 1)^2}}{d(1 + e^{ik_0(s_1+s_2)})} k_0 \tag{6}$$

By comparing the equations (5) and (6), we can find that the dispersion relation expressions of the surface modes excited by the equivalent anisotropic water layer and the SPRG are exactly the same, which analytically proves that this equivalence is feasible. Then we bring the equivalent anisotropic water layer into the commercial software COMSOL Multiphysics for verification.

The simulation results are shown in Fig. 2b. The area of the equivalent anisotropic water layer in Fig. 2b is within $-17m < y < -6m$, the wave source is at (0, 0), in order to satisfy the condition that surface waves appear on the metagrating: the wavelength is great than 4 times of groove depth of the metagrating ($\lambda > 4(s_1 + s_2)$), we set the wavelength around $46.4m$, this belongs to the case where the equivalent anisotropic water layer is located in the region of $y < y_{source}$, equation (3) should take the plus sign, from the field patterns in Fig. 2b, it can be seen that in the equivalent anisotropic water layer, the vertical displacement of water wave in the left part ($k_x < 0$) is weakened, while in the right part ($k_x > 0$) is enhanced, forming a unidirectional surface water wave propagating to the right, which is in good agreement with the theory. Then we expand the vortex water wave in Fig. 2b by plane waves, and solve the expression of water wave in each region analytically, and plot the field pattern of water wave corresponding to Fig. 2b, as shown in Fig. 2c (the derivation is in the Supplementary Materials Section One). In order to distinguish from Fig. 2b, the anisotropic water layer in Fig. 2c is marked with a solid red line. By comparing Fig. 2b and Fig. 2c, we can find that the field patterns in Fig. 2b and Fig. 2c are almost identical, which also proves that our numerical simulation and analytical calculation are in good agreement with each other. With this equivalent anisotropic water layer, we can simulate an infinitely long SPRG. We can use the method of splicing the equivalent anisotropic water layer and the perfectly matched layer (PML) to the two sides of the SPRG to extend it infinitely (which is directly called splicing model later), and then the numerical simulation of the SPRG can be carried out accurately. The simulation results are shown in Fig. 2d, where the area of the spliced model of the SPRG is $-17m < y < -6m$, the wave source is located at (0,0), and the wavelength is $47m$. The simulation results show that Fig. 2c and Fig. 2d are highly similar. The wave source excites the surface water wave propagating to the right in the SPRG, and then enters the equivalent water layer and continues to propagate to PML. The vertical displacement of the water wave in the right part is enhanced, while in the left part is greatly weakened, thereby forming a unidirectional surface water wave, which proves that the SPRG can indeed excite a unidirectional surface water wave. In order to make our equivalent model closer to the real Luoyang Bridge, we introduce the anisotropic gradient water layer into the equivalent model. Then we can write down a possible equivalent parameter of the equivalent gradient anisotropic water layer of the whole Luoyang Bridge by the layered method as follows (the derivation is in the Supplementary Materials Section One):

$$\overleftrightarrow{u}, g = \begin{cases} u_x = 0, u_y = \frac{(d+y+s_1+1)}{d} * u_0, g = \frac{d}{(d+y+s_1+1)} * g_0 & (-11m < y < -6m) \\ u_x = 0, u_y = \frac{a}{d} * u_0, g = \frac{d}{a} * g_0 & (-17m < y < -11m) \end{cases} \quad (7)$$

where $u_0 = tank(k_0 h_0)/k_0$, $k_0 = 2\pi/\lambda$, the background water depth of real Luoyang Bridge is $h_0 = 3m$, the wavelength $\lambda = 46.4m$.

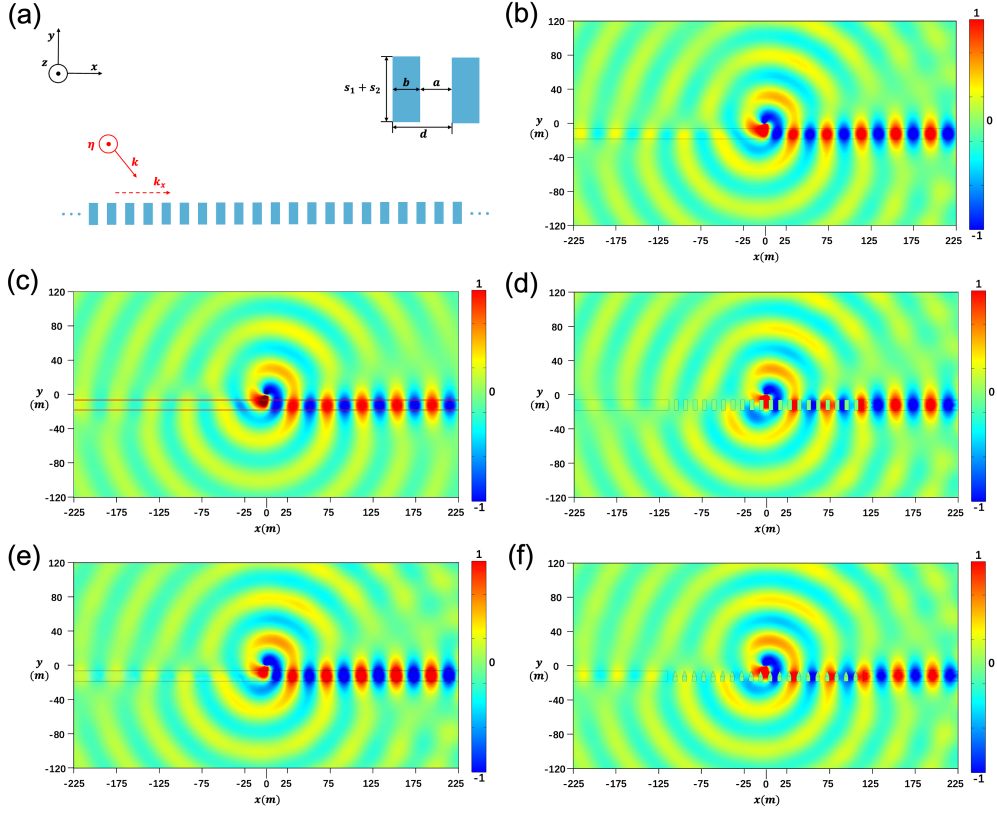

**FIG. 2.** Water wave field patterns of four equivalent models of Luoyang Bridge. (a) Schematic diagram of SPRG. The simulation (b) and the analytical solution (c) of filed pattern of water wave excited by the equivalent anisotropic water layer of the SPRG. (d) The simulated field pattern of water wave excited by the spliced model of the SPRG. The simulated field pattern of water wave excited by the equivalent anisotropic gradient water layer (e) and the spliced model (f) of Luoyang Bridge.

The simulation result of equivalent anisotropic gradient water layer is shown in Fig. 2e, in which the area of $-17m < y < -6m$ is the equivalent anisotropic gradient water layer, and the wave source is located at (0m,0m) with wavelength $\lambda = 46.4m$, which belongs to the case where the equivalent gradient water layer is located in the region of $y < y_{source}$. From the field pattern in

Fig. 2e, it can be seen that the vertical displacement of water wave in the left part ($k_x < 0$) in the equivalent gradient water layer is weakened, while in the right part ($k_x > 0$) is enhanced to form a unidirectional surface water wave propagating to the right. We extend the metagrating of Luoyang Bridge infinitely by the splicing method. Then we simulate and verify the splicing model, and the simulation result is shown in Fig. 2f. The area of $-17m < y < -6m$ in Fig. 2f is the equivalent gradient water layer and the splicing model of metagrating, where the wave source is located at (0m,0m), and the wavelength is still $\lambda = 46.4m$. From the simulation results, it can be seen that the Fig. 2e and Fig. 2f are highly similar, also forming a unidirectional propagation surface water wave, which proves that the metagrating can indeed excite a unidirectional propagation surface wave for water waves.

***Experimental Verifications.*** We then do some experiments to qualitatively observe the unidirectional propagation water wave excited near the metagrating. We print out a reduced metagrating with a 3D printer (Reduced by 100 times compared to Fig. 1b), its structure diagram is shown in Supplementary Figure 6c and its material is PLA plastic, and its structural dimensions are $a = 0.6\,cm$, $b = 0.5\,cm$, $d = 1.1\,cm$, $s_1 = 0.5cm$, $s_2 = 0.6cm$. The water depth in the experiment is $8\,mm$. The equipment used in the experiment is shown in Supplementary Figure 6a, b. A miniature propeller is used to excite the vortex water wave, we set its frequency $f$ to 5.1 Hz, according to the dispersion relation of water waves (2), we can find that the corresponding wavelength $\lambda$ is 4.64cm, the impermeable rigid body is a reduced metagrating (in cyan), and the black sponges on the left and right sides are used to eliminate the reflection of water waves at outer boundaries (detailed structure could be seen in Supplementary Materials Section Two).

The experimental results are shown in Fig. 3. Figure 3a is a schematic diagram of the observation of experimental results. The light shines from above and projects the trajectory of the water waves onto the bottom screen to facilitate our observation. Therefore, the experimental results below are all projections of water waves. In order to highlight the role of metagratings, we designed a set of control experiments as shown in Fig. 3b. The sample of the control group is in the upper part, which a flat plate without any structure. The sample of the experimental group is in the lower part, which is a reduced metagrating.

Due to the capillary effect of water, the projected area of the metagrating is larger than itself, making it difficult to observe the near-field surface water waves formed by vortex waves near the metagrating. However, when we carefully observe the simulation results in Fig. 2, we will find that obvious unidirectional phenomena can also be seen in the far field of 1 to 2 wavelengths near the wave source. This is because the field near the wave source is strong. Although the unidirectional surface water wave is evanescent wave, it can still propagate a certain distance. Therefore, we verify our conclusion by observing the unidirectional phenomenon in the far field near the wave source in the experiment.

Fig. 3c shows the experimental results without metagrating. It can be seen that a complete vortex water wave (a counterclockwise rotating propeller) has the same vertical displacement on both sides of the water wave with a wavelength of 4.64 cm (with frequency 5.1 Hz).

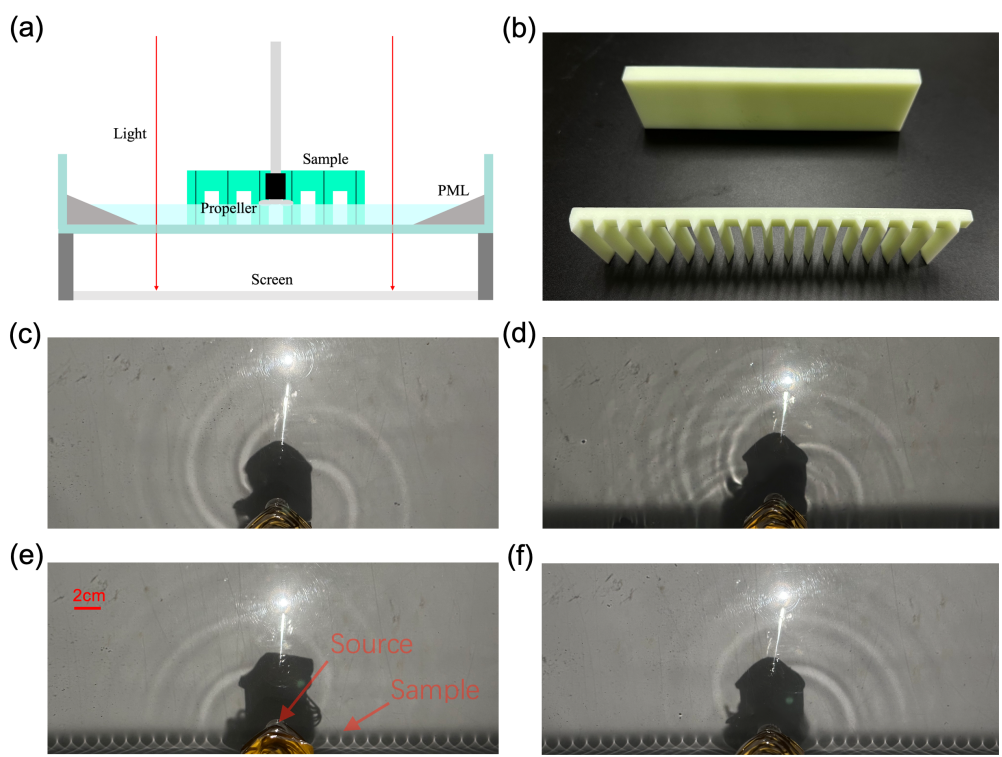

**FIG. 3.** Experimental results. (a) Side view of the experimental equipment. (b) The Sample of control (up) and experimental (down) group. (c) Vortex water wave excited by a rotating miniature propeller, the propeller is counterclockwise rotating. (d) Field pattern for the case when the rotating propeller is close (0.6 cm) to the flat plate. The propeller is counterclockwise rotating. (e) (f) Field pattern for the case when the rotating propeller is close (0.6 cm) to the reduced metagrating. The propeller is (clockwise) counterclockwise rotating.

Figure 3d shows the experimental results of the control group, The wavelength is still 4.64 cm, and the wave source (a counterclockwise rotating propeller) is put 0.6 cm away from the flat plate. It can be seen that the water wave on the left side of the wave source is strongly reflected after encountering the flat plate, and the vertical displacement of the water wave does not weaken at all.

Then we replaced the flat plate with reduced metagrating, the experimental results are shown in Fig. 3e and f. When the propeller rotates clockwise (counterclockwise), the vertical displacement of the water wave on the right (left) side of the wave source is weakened, forming a unidirectional surface water wave propagates to the left (right). At the same time, by comparing Fig. 3d and f, we can see that compared with the flat plate, the metagrating can significantly weaken the vertical displacement of the water wave on the left side of the wave source, proving that the metagrating can indeed excite unidirectional propagating surface water waves, which is in good agreement with our theory. (Fig. 3e and f have corresponding videos in Supplementary Materials Section Two)

We also conduct experiments near the real Luoyang Bridge. However, Luoyang Bridge is a protected area, we cannot conduct more systematic experiments, some unidirectional effects can also be seen (the experimental results are in the Supplementary Materials Section Two).

***Conclusion.*** We found that metagratings can also control water waves, the vortex water waves can excite unidirectional water waves near metagrating. At the same time, we calculate the equivalent anisotropic water layer model of metagrating. We verified the equivalent anisotropic water layer from analytical calculations and numerical simulations. Then we carried out experiments on a reduced metagrating and qualitatively observed the results similar to the theory. Our work provides a new method for controlling water waves, namely metagratings, which can help better understand the unidirectional propagation phenomena and physical mechanisms contained in metagratings and even in ancient building like Luoyang Bridge. It also has great significance to water wave metagrating design and the development and protection of ancient bridges.

## Supplementary Materials

### Section One: Theoretical Analysis

**(1) Corresponding relationship between electromagnetic wave equation for TM mode and general water wave equation.**

In our previous work, we found the corresponding relationship between electromagnetic wave for TM mode (Eq. (S1)) and shallow water wave equation ($\lambda \gg h$) (Eq. (S2)):

$$\nabla \cdot \left( \frac{\overleftrightarrow{\varepsilon}}{\det(\overleftrightarrow{\varepsilon})\,\varepsilon_0} \nabla H_z \right) + \mu\mu_0 \omega^2 H_z = 0 \tag{S1}$$

where $\overleftrightarrow{\varepsilon} = \begin{bmatrix} \varepsilon_x & 0 \\ 0 & \varepsilon_y \end{bmatrix}$, $\frac{\overleftrightarrow{\varepsilon}}{\det(\overleftrightarrow{\varepsilon})} = \begin{bmatrix} 1/\varepsilon_y & 0 \\ 0 & 1/\varepsilon_x \end{bmatrix}$, $\varepsilon_0$ is the dielectric constant, $\overleftrightarrow{\varepsilon}$ is the relative dielectric constant tensor, $\mu_0$ is permeability constant, $\mu$ is the relative permeability, $\omega$ is the angular frequency;

$$\nabla \cdot (\overleftrightarrow{h} \nabla p) + \frac{\omega^2}{g} p = 0 \tag{S2}$$

where $\overleftrightarrow{h}$ is a tensor with $\overleftrightarrow{h} = \begin{bmatrix} h_x & 0 \\ 0 & h_y \end{bmatrix}$, $p$ is the hydrostatic pressure of water surface ($p = \rho g \eta$), $\rho$ is the fluid density, $g$ is the gravitational acceleration, $\eta$ is the vertical displacement of the water wave (as shown in the lower right corner of Fig. 1(b) in the main text), $\omega$ is the angular frequency.

Then we compare equations (S1) and (S2), we find that:

$$H_z \leftrightarrow p, \frac{1}{\varepsilon_y \varepsilon_0} \leftrightarrow \frac{h_x}{\rho}, \frac{1}{\varepsilon_x \varepsilon_0} \leftrightarrow \frac{h_y}{\rho}, \frac{1}{\mu\mu_0} \leftrightarrow \rho g \tag{S3}$$

Through Eq. (S3), we can manipulate the movement of water waves by adjusting water depth $h$ and gravitational acceleration $g$, just as we manipulate electromagnetic waves by adjusting the dielectric constant $\varepsilon$ and permeability $\mu$. Let us now see whether this corresponding relationship still exist in water waves with a general water depth ($\lambda \sim h$).

We write the general water equation in anisotropic water layers [43]:

$$\nabla \cdot \left( \frac{\tanh(k\overleftrightarrow{h})}{k} \cdot \nabla p \right) + \frac{\omega^2}{g} \eta = 0 \tag{S4}$$

the water wave equation is governed by nonlinear dispersion $\omega = \sqrt{g \tanh(kh) k}$. $k$ is propagation wave number of water wave.

Through Eq. (S4), we can find that when the water depth doesn't meet the condition of shallow water ($\lambda \gg h$), the control equation of water wave becomes nonlinear, which make it difficult to correspond with the electromagnetic waves.

Here we introduce a new variable $u$ (equivalent depth), $u = tank(kh)/k$, so $\overleftrightarrow{u} = tank(k\overleftrightarrow{h})/k \Rightarrow \begin{bmatrix} u_x & 0 \\ 0 & u_y \end{bmatrix} = \begin{bmatrix} \tanh(kh_x)/k & 0 \\ 0 & \tanh(kh_y)/k \end{bmatrix}$. In this way, we can make a simplification of equation (S2):

$$\nabla \cdot (\overleftrightarrow{u} \cdot \nabla p) + \frac{\omega^2}{g}\eta = 0 \tag{S5}$$

And the dispersion become $\omega = \sqrt{guk}$.

Then we get a new corresponding relationship:

$$H_z \leftrightarrow p, \frac{1}{\varepsilon_y \varepsilon_0} \leftrightarrow \frac{u_x}{\rho}, \frac{1}{\varepsilon_x \varepsilon_0} \leftrightarrow \frac{u_y}{\rho}, \frac{1}{\mu\mu_0} \leftrightarrow \rho g \tag{S6}$$

Therefore, the movement of water wave with a general water depth can still be manipulated by adjusting water depth $h$ and gravitational acceleration $g$, even in the existence of nonlinearity, yet the water depth $h$ needs some adjustment according to the wave number $k$.

With these corresponding relationships, we can solve some manipulation parameters of metagratings for water waves.

**(2) Dispersion relationship of the surface mode of SPRG (corresponding to equation (7) in the main text)**

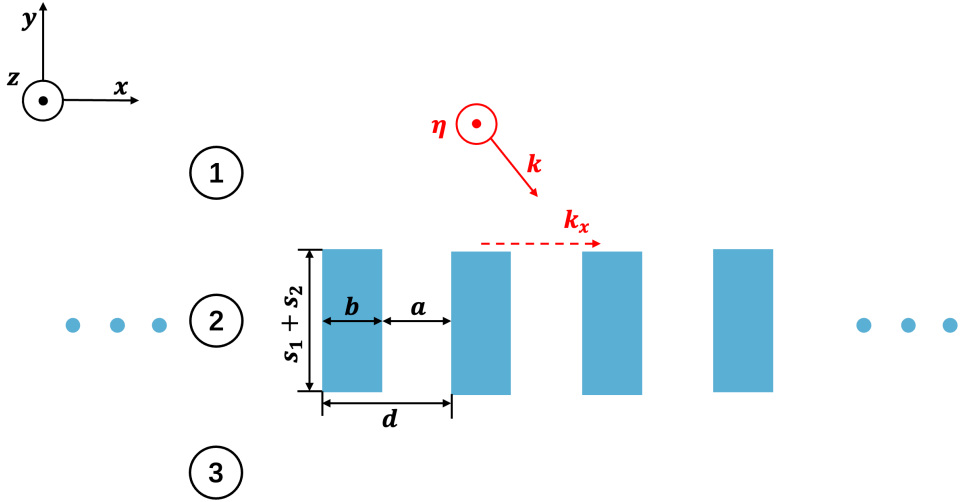

**Supplementary Figure 1.** SPRG structure diagram. The length of a single cycle unit is $s_1 + s_2$, the width is $b$, the interval between two cycle units is $a$ and the cycle period length is $d$. The incident wave is water wave, propagates along the $x$ direction, the vertical displacement of water wave $\eta$ is in the $z$ direction, and propagation direction of water wave $k$ is in the $x - y$ plane.

Because the location of the SPRG does not affect its transmission and reflection coefficient, for the convenience of solving, we put the upper bound of SPRG at $y = 0$. In Region 1 of Supplementary Figure 1, the total vertical displacement of water wave can be written as the superposition of the vertical displacement of incident water wave and the vertical displacement of water wave reflected of all diffraction orders [47,38]:

$$\eta_1 = e^{ik_x x} e^{-ik_y y} + \sum_{-\infty}^{\infty} R_n e^{ik_x^n x} e^{ik_y^n y} \tag{S7}$$

where $k_x^n = k_x + 2\pi n/d$ and $k_y^n = \sqrt{k_0^2 - (k_x^n)^2}$, $R_n$ is the reflection coefficient of the $n$th diffraction order. As the period length $d$ of the array is much smaller than the incident wavelength, all diffraction orders can be ignored except the specular reflection order. Therefore, the vertical displacement of water wave of Region 1 can be written as:

$$\eta_1 = e^{ik_x x} e^{-ik_y y} + R e^{ik_x x} e^{ik_y y} \tag{S8}$$

where $R$ is the specular reflection coefficient.

In Region 2, since the interval between two cycle units $a$ is much smaller than the incident wavelength, there is only the fundamental mode of the waveguide in the slit ($|x - x_j| < a/2$, where $x_j$ is the middle coordinate of each period), so the total vertical displacement of water wave in the slit of Region 2 is [47]:

$$\eta_2 = c_1 e^{-ik_0 y} + c_2 e^{ik_0 y} \quad (|x - x_j| < a/2) \tag{S9}$$

where $c_1$ is the amplitude coefficient of the forward propagating water wave (along the negative $y$-axis) and $c_2$ is the amplitude coefficient of the backward propagating water wave (along the positive $y$-axis).

In Region 3, the vertical displacement of water wave can be written in the form of transmitted waves, so the vertical displacement of water wave can be written as [38]:

$$\eta_3 = T e^{ik_x x} e^{-ik_y y} \tag{S10}$$

where $T$ is the transmission coefficient.

We assume the flow of water wave is $\vec{\zeta}$, then $\vec{\zeta} = \frac{igh}{\omega}(\nabla \eta) \Rightarrow \zeta_y = \frac{igh}{\omega}\frac{\partial \eta}{\partial y}$ [48], where $h$ is the water depth, $g$ is the gravitational acceleration. The background water depth and gravitational acceleration in Supplementary Figure 1 is $h_0$ and $g_0$, then we can get the expression for the total flow of the water wave in the $y$ direction in Region 1 as:

$$\zeta_{1y} = \left(\frac{-ig_0 h_0 k_y}{\omega}\right) * \left(e^{ik_x x} e^{-ik_y y} - R e^{ik_x x} e^{ik_y y}\right) \tag{S11}$$

In the Region 2, when $|x - x_j| < a/2$ (where $x_j$ is the middle coordinate of each period), the flow of water wave in the $y$ direction can pass through it, so we can get:

$$\zeta_{2y} = \left(\frac{-ig_0 h_0 k_0}{\omega}\right) * \left(c_1 e^{-ik_0 y} - c_2 e^{ik_0 y}\right) \tag{S12}$$

When $a/2 < |x - x_j| < d/2$, the SPRG is a rigid body, and the flow of water wave in the $y$ direction cannot pass through it. Therefore, $\zeta_{2y} = \frac{ig_0 h_0}{\omega}\frac{\partial \eta_2}{\partial y} = 0$, while the total flow of the water wave in the $y$ direction of Region 2 is:

$$\zeta_{2y} = \begin{cases} \left(\frac{-ig_0 h_0 k_0}{\omega}\right) * \left(be^{-ik_0 y} - ce^{ik_0 y}\right) & (|x - x_j| < a/2) \\ 0 & (a/2 < |x - x_j| < d/2) \end{cases} \quad (S13)$$

The flow of the water wave in the $y$ direction of Region 3 is:

$$\eta_3 = \left(\frac{-ig_0 h_0 k_y}{\omega}\right) * Te^{ik_x x} e^{-ik_y y} \quad (S14)$$

At $y = 0$:

$$\eta_1 = (1 + R)e^{ik_x x} \quad (S15.1)$$

$$\eta_2 = c_1 + c_2 \quad (|x - x_j| < a/2) \quad (S15.2)$$

$$\zeta_{1y} = \left(\frac{-ig_0 h_0 k_y}{\omega}\right) * (1 - R)e^{ik_x x} \quad (S15.3)$$

$$\zeta_{2y} = \begin{cases} \left(\frac{-ig_0 h_0 k_0}{\omega}\right) * (c_1 - c_2) & (|x - x_j| < a/2) \\ 0 & (a/2 < |x - x_j| < d/2) \end{cases} \quad (S15.4)$$

At $y = -(s_1 + s_2)$:

$$\eta_2 = c_1 e^{ik_0(s_1+s_2)} + c_2 e^{-ik_0(s_1+s_2)} \quad (|x - x_j| < a/2) \quad (S16.1)$$

$$\eta_3 = Te^{ik_y(s_1+s_2)} e^{ik_x x} \quad (S16.2)$$

$$\zeta_{2y} = \begin{cases} \left(\frac{-ig_0 h_0 k_0}{\omega}\right) * \left(be^{ik_0(s_1+s_2)} - ce^{-ik_0(s_1+s_2)}\right) & (|x - x_j| < a/2) \\ 0 & (a/2 < |x - x_j| < d/2) \end{cases} \quad (S16.3)$$

$$\zeta_{3y} = \left(\frac{-ig_0 h_0 k_y}{\omega}\right) * Te^{ik_y(s_1+s_2)} e^{ik_x x} \quad (S16.4)$$

At $y = 0$, when $|x - x_j| < a/2$, surface hydrostatic pressure $p$ ($p = \rho g \eta$, $\rho$ is fluid density) should be continuous (for the detailed derivation process, please refer to the Supplementary Materials of reference[43]), so $p_1 = p_2$. When $|x - x_j| < d/2$, $\zeta_y$ should be continuous, so $\zeta_{1y} = \zeta_{2y}$, namely:

$$(1 + R)e^{ik_x x} = (c_1 + c_2) \quad (|x - x_j| < a/2) \quad (S17)$$

$$k_y * (1 - R)e^{ik_x x} = \begin{cases} k_0 * (c_1 - c_2) & (|x - x_j| < a/2) \\ 0 & (a/2 < |x - x_j| < d/2) \end{cases} \quad (S18)$$

At $y = -(s_1 + s_2)$, when $|x - x_j| < a/2$, surface hydrostatic pressure $p$ should be continuous, so $p_2 = p_3$. When $|x - x_j| < d/2$, $\zeta_y$ should be continuous, so $\zeta_{2y} = \zeta_{3y}$, namely:

$$Te^{ik_y(s_1+s_2)} e^{ik_x x} = \left(c_1 e^{ik_0(s_1+s_2)} + c_2 e^{-ik_0(s_1+s_2)}\right) \quad (|x - x_j| < a/2) \quad (S19)$$

$$k_y * Te^{ik_y(s_1+s_2)}e^{ik_x x} = \begin{cases} k_0 * \left(c_1 e^{ik_0(s_1+s_2)} - c_2 e^{-ik_0(s_1+s_2)}\right) & (|x-x_j| < a/2) \\ 0 & (a/2 < |x-x_j| < d/2) \end{cases} \quad (S20)$$

Integrate (S11) (S13) on $|x-x_j| < a/2$ and (S19) (S20) on $a/2 < |x-x_j| < d/2$ to obtain:

$$a(1+R) = (b+c)s_0 \quad (S21)$$

$$d\frac{k_y}{k_0}(1-R) = (c_1 - c_2)s_0 \quad (S22)$$

$$aTe^{ik_y(s_1+s_2)} = \left(c_1 e^{ik_0(s_1+s_2)} + c_2 e^{-ik_0(s_1+s_2)}\right)s_0 \quad (S23)$$

$$d\frac{k_y}{k_0}Te^{ik_y(s_1+s_2)} = \left(c_1 e^{ik_0(s_1+s_2)} - c_2 e^{-ik_0(s_1+s_2)}\right)s_0 \quad (S24)$$

where $s_0 = \int_{x_j-a/2}^{x_j+a/2} e^{-ik_x x} dx = e^{-ik_x x_j}\frac{a*\sin(k_x a/2)}{k_x a/2} = e^{-ik_x x_j} sinc(k_x a/2)$.

From equations (S21)~(S24) we obtain:

$$R = \frac{\left(e^{i2k_0(s_1+s_2)} - 1\right)\left(a^2 k_0^2 - d^2 k_y^2\right)}{\left(ak_0 + dk_y\right)^2 - \left(ak_0 e^{ik_0(s_1+s_2)} - dk_y e^{ik_0(s_1+s_2)}\right)^2} \quad (S25)$$

Then by extending (S25) to $k_x > k_0 (k_y = i\sqrt{k_x^2 - k_0^2})$, and calculating the zero point of the denominator [49], we obtain the dispersion relationship of the surface mode of SPRG:

$$k_x = \frac{\sqrt{d^2(e^{ik_0(s_1+s_2)}+1)^2 - a^2(e^{ik_0(s_1+s_2)}-1)^2}}{d(1+e^{ik_0(s_1+s_2)})} k_0 \quad (S26)$$

which is the equation (5) in the main text.

### (3) Reflection coefficient and dispersion relationship of the surface mode of the equivalent anisotropic water layer (corresponding to equation (8) in the main text)

We first write the revised water equation [43] in anisotropic water layers:

$$\nabla \cdot \left(\frac{\tanh(k\overleftrightarrow{h})}{k} \cdot \nabla p\right) + \frac{\omega^2}{g}\eta = 0 \quad (S27)$$

where $\overleftrightarrow{h}$ is a tensor with $\overleftrightarrow{h} = \begin{bmatrix} h_x & 0 \\ 0 & h_y \end{bmatrix}$, $p$ is the hydrostatic pressure of water surface ($p = \rho g \eta$), $\rho$ is the fluid density, $g$ is the gravitational acceleration, $\eta$ is the vertical displacement of the water wave (as shown in the lower right corner of Fig. 1(B) in the main text), $\omega$ is the angular frequency, the surface wave equation above the water wave is governed by nonlinear dispersion $\omega = \sqrt{g\tanh(kh)}k$. $k$ is propagation wave number of water wave.

We define a new variable $u$ (reduced water depth), $u = tanh(kh)/k$, so $\overleftrightarrow{u} = tank(k\overleftrightarrow{h})/k = \begin{bmatrix} u_x & 0 \\ 0 & u_y \end{bmatrix}$. In this way, we can make a simplification of equation (S27):

$$\nabla \cdot (\overleftrightarrow{u} \cdot \nabla p) + \frac{\omega^2}{g}\eta = 0 \qquad (S28)$$

And the dispersion becomes $\omega = \sqrt{guk}$.

With the equation (S28), we can write the equivalent water layer with an anisotropic depth and a specific gravity of the SPRG, as shown in Supplementary Figure 2. The parameters of the equivalent water layer are shown below:

$$u_{2x} = 0, \; u_{2y} = a/d * u_0, \; g_2 = d/a * g_0 \qquad (S29)$$

where $u_0 = \tanh(kh_0)/k$, $h_0$ is the water depth of Region 1 and Region 3, $g_0$ is the gravitational acceleration of Region 1 and Region 3 in Supplementary Figure 2.

We assume that the reduced water depth and gravitational acceleration of Region 1 is $u_1 = u_0, g_1 = g_0$, the water depth and gravitational acceleration of Region 2 is $\overleftrightarrow{u_2} = \begin{bmatrix} u_{2x} & 0 \\ 0 & u_{2y} \end{bmatrix} = \begin{bmatrix} 0 & 0 \\ 0 & a/d*u_0 \end{bmatrix}, g_2 = d/a * g_0$, the water depth and gravitational acceleration of Region 3 is $u_3 = u_0, g_3 = g_0$, $\theta_1$ is the angle between the direction of the incident water wave and the normal direction of the interface, $\theta_2$ is the angle between the direction of the refraction water wave and the normal direction of the interface as shown in the Supplementary Figure 3.

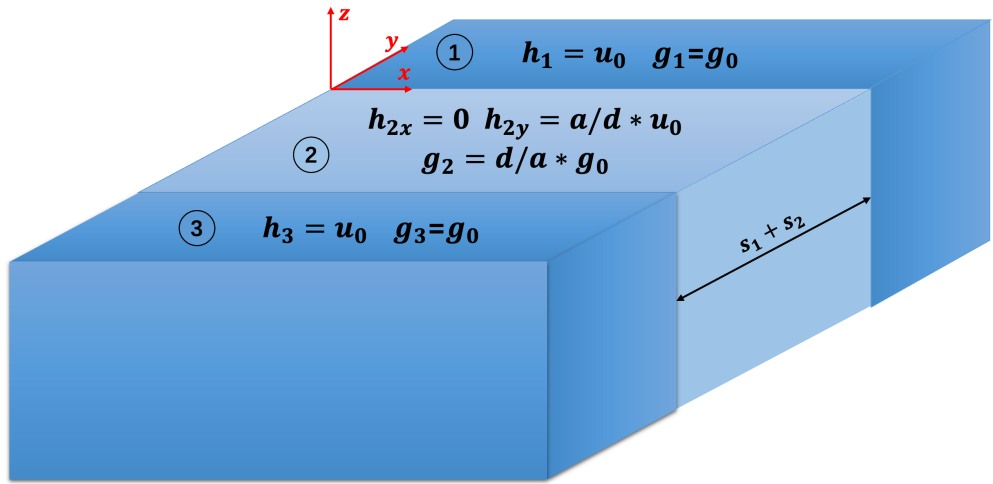

**Supplementary Figure 2.** Anisotropic water layer equivalent model.

When the water wave goes from Region 1 to Region 2, the angular frequency $\omega$ remains the same, so $\omega_2 = \omega_1 = \sqrt{g_0 u_0} k_0$, $k_0$ is water wave propagation wavenumber of Region 1.

Through equation (S28), we can find the relationship between the wave vector components in each direction in the equivalent anisotropic water layer (Region 2):

$$\frac{g_2 u_{2x}}{g_0 u_0} k_{2x}^2 + \frac{g_2 u_{2y}}{g_0 u_0} k_{2y}^2 = k_0^2 \tag{S30}$$

Then we can get:

$$k_{2x} = \frac{k_0 \sin\theta_2}{\sqrt{g_2 u_{2x}/(g_0 h_0)}} \tag{S31}$$

$$k_{2y} = \frac{k_0 \cos\theta_2}{\sqrt{g_2 u_{2y}/(g_0 h_0)}} \tag{S32}$$

where $\theta_1$ is the angle between the direction of the incident water wave and the normal direction of the interface, $\theta_2$ is the angle between the direction of the refraction water wave and the normal direction of the interface as shown in the Supplementary Figure 3.

Conservation of the parallel wave vector components at the interface of Region 1 and Region 2 leads to $k_{1x} = k_{2x}$, then we can get:

$$k_0 \sin\theta_1 = \frac{k_0 \sin\theta_2}{\sqrt{g_2 u_{2x}/(g_0 u_0)}} \tag{S33}$$

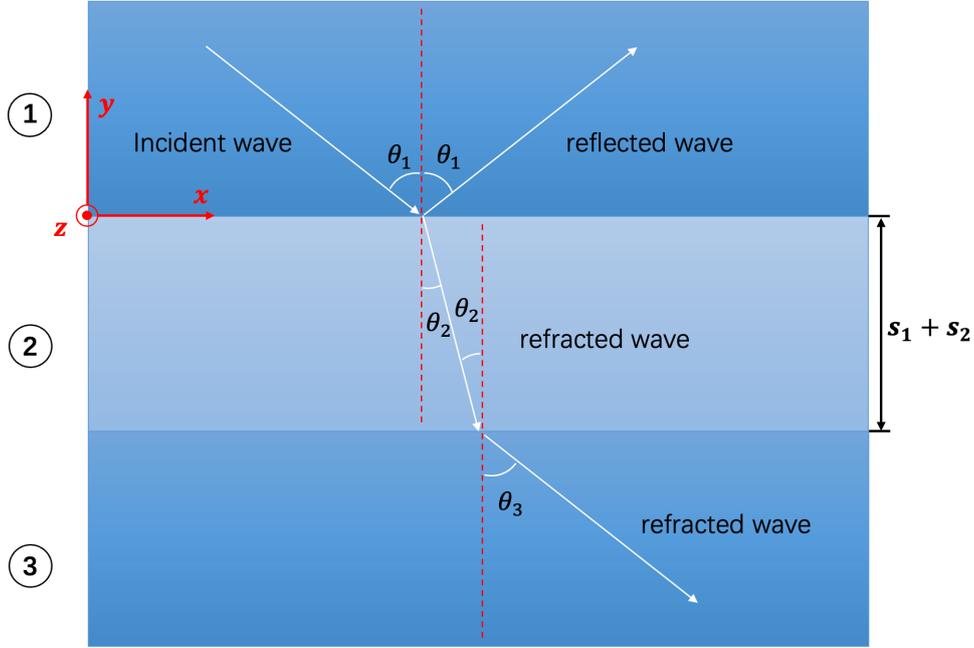

**Supplementary Figure 3.** Propagation of water waves in different regions.

which leads to:

$$k_{2y} = \frac{k_0 \cos\theta_2}{\sqrt{g_2 u_{2y}/(g_0 u_0)}} \tag{S34}$$

We bring $u_{2x} = 0, u_{2y} = a/d * u_0, g_2 = d/a * g_0$ into $k_0 \sin\theta_1 = \frac{k_0 \sin\theta_2}{\sqrt{g_2 u_{2y}/(g_0 u_0)}}$ to get $\theta_2 = \arcsin(\sqrt{g_2 u_{2y}/(g_0 u_0)} * \sin\theta_1) \approx 0$, so $k_{2y} = \frac{k_0 \cos\theta_2}{\sqrt{g_2 u_{2y}/(g_0 u_0)}} \approx k_0$.

Therefore, no matter what angle the incident water wave enters into the anisotropic equivalent water layer, the refraction angle will become 0 degree, and the wave number propagating in the $y$ direction will become $k_0$.

At the same time, $k_0 sin\theta_1 = \frac{k_0 sin\theta_2}{\sqrt{g_2 u_{2x}/(g_0 h_0)}} = \frac{k_0 sin\theta_3}{\sqrt{g_3 u_3/(g_0 h_0)}}$, because the water depth and gravity are the same in Region 1 and Region 3, so $k_0 sin\theta_1 = k_0 sin\theta_3$, i.e., $\theta_1 = \theta_3$, $k_{3y} = k_{1y}$. Then we can use the transmission matrix method [S50] to calculate the reflection coefficient of the surface of Region 1.

From the transmission matrix theory, we can obtain the relationship between the amplitude of the incident wave and the outgoing wave as:

$$\begin{bmatrix} b_1 \\ c_1 \end{bmatrix} = D_{1\to 2} P_{2\to 3} D_{2\to 3} \begin{bmatrix} b_3 \\ c_3 \end{bmatrix} = M \begin{bmatrix} b_3 \\ c_3 \end{bmatrix} \quad (S35)$$

where $b_n$ is the amplitude coefficient of the forward propagating water wave (along the negative $y$-axis) in the Region $n$ and $c_n$ is the amplitude coefficient of the backward propagating water wave (along the positive $y$-axis) in the Region $n (n = 1,2,3)$.

$$D_{1\to 2} = \frac{1}{2}\begin{bmatrix} 1+\rho_1 & 1-\rho_1 \\ 1-\rho_1 & 1+\rho_1 \end{bmatrix}, P_{2\to 3} = \begin{bmatrix} e^{-ik_{2y}(s_1+s_2)} & 0 \\ 0 & e^{ik_{2y}(s_1+s_2)} \end{bmatrix},$$

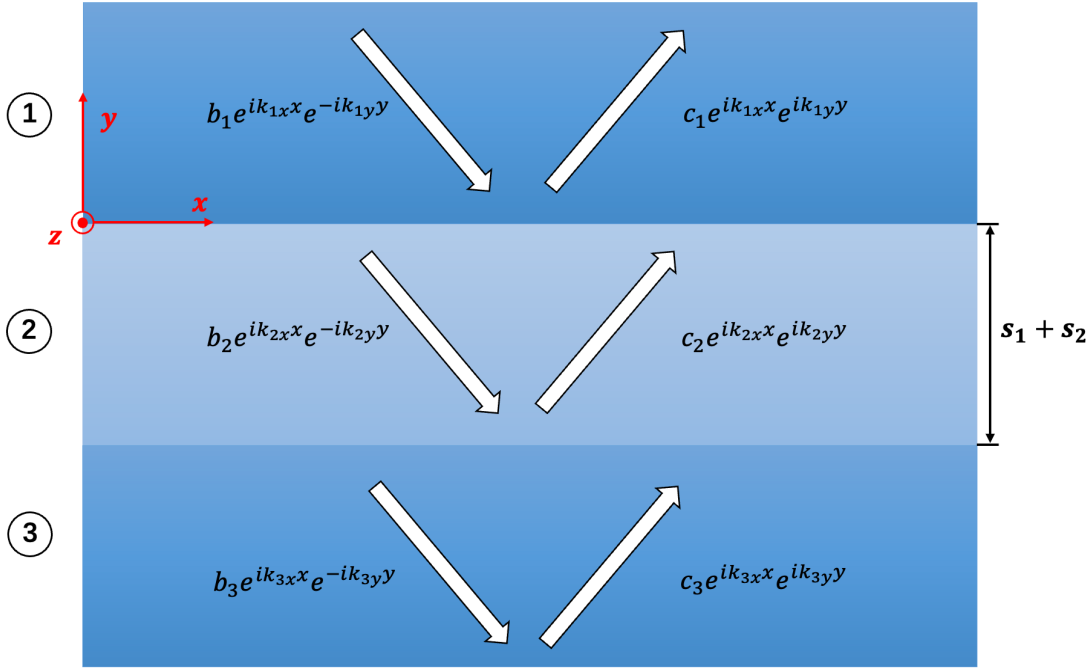

**Supplementary Figure 4.** Schematic diagram of water wave propagation in the equivalent water layer. The incident wave is water wave, which propagates obliquely along the $x$ direction. The vertical displacement of water wave $\eta$ is in the $z$ direction, and the propagation direction of water wave $k$ is in the $x - y$ plane.

$D_{2\to 3} = \frac{1}{2}\begin{bmatrix} 1+\rho_2 & 1-\rho_2 \\ 1-\rho_2 & 1+\rho_2 \end{bmatrix}, \rho_1 = \frac{k_{2y}u_{2y}}{k_{1y}u_{1y}}, \rho_2 = \frac{k_{3y}u_{3y}}{k_{2y}u_{2y}}$. Then bringing $h_{3y} = h_{1y} = h_0, h_{2y} = a/d * h_0, k_{3y} = k_{1y}, k_{2y} \approx k_0$ into Eq. (S35) to get:

$$M = \begin{bmatrix} M_{11} & M_{12} \\ M_{21} & M_{22} \end{bmatrix} = \frac{1}{4}\begin{bmatrix} 1+\frac{ak_0}{dk_{1y}} & 1-\frac{ak_0}{dk_{1y}} \\ 1-\frac{ak_0}{dk_{1y}} & 1+\frac{ak_0\varepsilon_{1x}}{dk_{1y}\varepsilon_{2x}} \end{bmatrix}\begin{bmatrix} e^{-ik_0(s_1+s_2)} & 0 \\ 0 & e^{ik_0(s_1+s_2)} \end{bmatrix}\begin{bmatrix} 1+\frac{dk_{1y}}{ak_0} & 1-\frac{dk_{1y}}{ak_0} \\ 1-\frac{dk_{1y}}{ak_0} & 1+\frac{dk_{1y}}{ak_0} \end{bmatrix} \quad (S36)$$

Then we obtain:
$$M_{11} = \frac{1}{4}\left[\left(2 + \frac{dk_{1y}}{ak_0} + \frac{ak_0}{dk_{1y}}\right)e^{-ik_0(s_1+s_2)} + \left(2 - \frac{dk_{1y}}{ak_0} - \frac{ak_0}{dk_{1y}}\right)e^{ik_0(s_1+s_2)}\right]$$

$$M_{12} = \frac{1}{4}\left[\left(\frac{ak_0}{dk_{1y}} - \frac{dk_{1y}}{ak_0}\right)e^{-ik_0(s_1+s_2)} + \left(\frac{dk_{1y}}{ak_0} - \frac{ak_0}{dk_{1y}}\right)e^{ik_0(s_1+s_2)}\right]$$

$$M_{21} = \frac{1}{4}\left[\left(\frac{dk_{1y}}{ak_0} - \frac{ak_0}{dk_{1y}}\right)e^{-ik_0(s_1+s_2)} + \left(\frac{ak_0}{dk_{1y}} - \frac{dk_{1y}}{ak_0}\right)e^{ik_0(s_1+s_2)}\right]$$

$$M_{22} = \frac{1}{4}\left[\left(2 - \frac{dk_{1y}}{ak_0} - \frac{ak_0}{dk_{1y}}\right)e^{-ik_0(s_1+s_2)} + \left(2 + \frac{dk_{1y}}{ak_0} + \frac{ak_0}{dk_{1y}}\right)e^{ik_0(s_1+s_2)}\right]$$

Eventually, we can get the reflection coefficient $R$ of Region 1 through the $M$ matrix.

$$R = \frac{M_{21}}{M_{11}} = \frac{\left(\frac{dk_{1y}}{ak_0} - \frac{ak_0}{dk_{1y}}\right)e^{-ik_0(s_1+s_2)} + \left(\frac{ak_0}{dk_{1y}} - \frac{dk_{1y}}{ak_0}\right)e^{ik_0(s_1+s_2)}}{\left(2 + \frac{dk_{1y}}{ak_0} + \frac{ak_0}{dk_{1y}}\right)e^{-ik_0(s_1+s_2)} + \left(2 - \frac{dk_{1y}}{ak_0} - \frac{ak_0}{dk_{1y}}\right)e^{ik_0(s_1+s_2)}} \quad (S37)$$

Then by extending Eq. (S37) to $k_x > k_0 (k_y = i\sqrt{k_x^2 - k_0^2})$, and calculating the zero point of the denominator [49], we obtain the dispersion relationship of the surface mode of equivalent anisotropic water layer:

$$k_x = \frac{\sqrt{d^2(e^{ik_0(s_1+s_2)} + 1)^2 - a^2(e^{ik_0(s_1+s_2)} - 1)^2}}{d(1 + e^{ik_0(s_1+s_2)})} k_0 \quad (S38)$$

By comparing the equations (S26) and (S38), we can find that the dispersion relationship of the surface mode of the equivalent anisotropic water layer and the SPRG are exactly the same, which proves that this equivalence is feasible. This is the equation (6) in the main text.

**(4) Analytical solution of field pattern in Fig. 2c in the main text**

The total vertical displacement of water wave with angular momentum is: $\eta = H_1(k_0 r)e^{i\theta}$, where $H_1$ is the Hankel function of the first kind, $k_0$ is wave vector. $k_0 = 2\pi/\lambda$, $\lambda$ is the wavelength. $r$ and $\theta$ are cylindrical coordinate systems. Because $H_1(k_0 r)e^{i\theta}$ is in the form of cylindrical waves. The equivalent anisotropic water layer is rectangular, so it is difficult to directly solve the expression of the field patterns excited by $H_1(k_0 r)e^{i\theta}$. Therefore, we first convert $H_1(k_0 r)e^{i\theta}$ into the form of plane wave by using the method of spatial Fourier transform [51], as shown in (S32):

$$\eta = H_1(k_0 r)e^{i\theta} = \int_{-\infty}^{\infty} \psi(k_x) e^{ik_x x} e^{ik_y|y-y_{source}|} dk_x \quad (S39)$$

where $\psi(k_x) = \begin{cases} \psi_1(k_x) = \frac{-i}{\pi k_0}\left[\frac{k_x + ik_y}{k_y}\right], y > y_{source} \\ \psi_2(k_x) = \frac{-i}{\pi k_0}\left[\frac{k_x - ik_y}{k_y}\right], y < y_{source} \end{cases}$, $y_{source}$ is the location of the excited source, $k_x$ is the wave number in the $x$ direction, $k_y = \sqrt{k_0^2 - k_x^2}$ is the wave number in the $y$ direction.

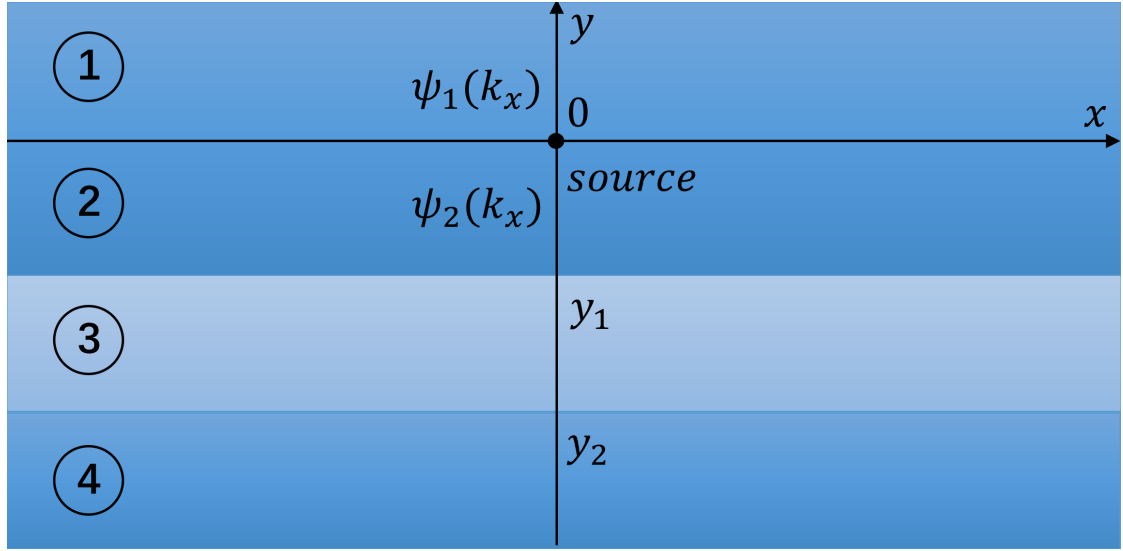

**Supplementary Figure 5.** Water waves with angular momentum are incident on anisotropic water layers.

For the convenience of calculation, we put the wave source at the origin, namely, $y_{source} = 0$. In this way, Supplementary Figure 5 is divided into four Regions, its Region 1 and 2 correspond to Region 1 in Supplementary Figure 3, and its Region 3 and 4 correspond to Region 2 and 3 in Supplementary Figure 3, so the reduced water depth of the Region 1,2,4 is $u_0$, the gravitational acceleration of the Region 1,2,4 is $g_0$. The relationship between the propagation wave numbers in the $x$ and $y$ directions of Region 1, 2, 4 is: $k_x^2 + k_{ny}^2 = k_0^2 (n = 1,2,4)$, the reduced water depth of Region 3 is $u_{3x} = 0, u_{3y} = a/d * u_0$, the gravitational acceleration of Region 3 is $g_3 = d/a * g_0$. The propagation wave number in the $x$ direction of the Region 3 does not change, and the wave number propagating in the $y$ direction of the Region 3 is $k_{3y} = k_0$. The calculation process of transmission and reflection has been written in details in the previous calculation sections. Here we briefly describe the calculation process and write the transmittance and reflection coefficient directly.

Using continuity of boundary conditions, we can solve for the transmission and reflection coefficients $R_{23}$ and $T_{23}$ at the junction of Region 2 and 3 and the transmission and reflection coefficients $R_{34}$ and $T_{34}$ at the junction of regions 2 and 3 as:

$$R_{23} = \frac{1-\alpha}{1+\alpha}, T_{23} = \frac{2}{1+\alpha} \qquad (S40)$$

$$R_{34} = \frac{\alpha-1}{\alpha+1}, T_{34} = \frac{2\alpha}{1+\alpha} \qquad (S41)$$

where $\alpha = ak_0/(d\sqrt{k_0^2 - k_x^2})$.

Then we can write the total transmission and reflection coefficient expression and water wave amplitude expression of each Region as follows [52]: The total water wave in Region 1 ($y > 0$) is formed by the superposition of the water wave incident from the origin of the wave source upward to Region 1 and the water wave reflected from Region 2, so the total vertical displacement of water wave of Region 1 is:

$$\eta_1 = \int_{-\infty}^{\infty} \left( \psi_1(k_x) e^{ik_{1y}y} + \psi_2(k_x) R e^{ik_{2y}2y_1} e^{ik_{1y}y} \right) e^{ik_x x} dk_x \quad (S42)$$

The total water wave in Region 2 ($y_1 < y < 0$) is formed by the superposition of the water wave incident from the origin of the wave source downward to Region 2 and the water wave reflected from the junction of Region 2 and 3, so the total vertical displacement of water wave of Region 2 is as follows:

$$\eta_2 = \int_{-\infty}^{\infty} \psi_2(k_x) \left( e^{-ik_{2y}y} + R e^{-ik_{2y}(2y_1 - y)} \right) e^{ik_x x} dk_x \quad (S43)$$

The total water wave in Region 3 ($y_2 < y < y_1$) is formed by the superposition of the water wave transmitted from Region 2 to Region 3 and the water wave reflected from the junction of Region 3 and 4, so the total vertical displacement of water wave of Region 3 is as follows:

$$\eta_3 = \int_{-\infty}^{\infty} \psi_2(k_x) A e^{-ik_{2y}y_1} \left( e^{-ik_0(y - y_1)} + R_{34} e^{-ik_0(2y_2 - y - y_1)} \right) e^{ik_x x} dk_x \quad (S44)$$

The total water wave in Region 4 ($y < y_2$) is formed by the water wave transmitted from Region 3 to Region 4, so the total vertical displacement of water wave of Region 4 is as follows:

$$\eta_4 = \int_{-\infty}^{\infty} \psi_2(k_x) T e^{-ik_{2y}y_1} e^{-ik_0(y_2 - y_1)} e^{-ik_{4y}(y - y_2)} e^{ik_x x} dk_x \quad (S45)$$

where

$$k_{ny}^2 = \sqrt{k_0^2 - k_x^2} \ (n = 1,2,4) \quad (S46)$$

$$R = P\left[ R_{23} + R_{34} e^{-i2k_0(y_2 - y_1)} \right] \quad (S47)$$

$$A = PT_{23}, \quad T = PT_{23}T_{34} \quad (S48)$$

$$P = \left[ 1 + R_{23} R_{34} e^{-i2k_0(y_2 - y_1)} \right]^{-1} \quad (S49)$$

Bringing (S40~S41) and (S46~S49) into (S42~S45) and then integrating them, we can plot the field pattern as shown in Fig. 2c in the main text.

**(5) Approximate water depth and gravitational acceleration of the equivalent gradient anisotropic water layer of the Luoyang Bridge (corresponding to the equation (6) in the main text)**

Because the single circle structure of Luoyang Bridge can be divided into an isosceles triangle and a rectangle, the equivalent gradient anisotropic water layer of Luoyang Bridge should also be divided into two regions. Here, by taking Fig. 2f in the main text as an example, we assume that the reduced water depth and gravitational acceleration of background region ($y > -6m$ and $y < -17m$) is $u_0$ and $g_0$, the rectangular part ($-17m < y < -11m$) under Luoyang Bridge has exactly the same structure as SPRG, so the equivalent water layer depth of the lower part is also $u_x = 0, u_y = a/d * u_0, g = d/a * g_0$. For the equivalent anisotropic gradient water layer in the upper isosceles triangular area ($-11m < y < -6m$), we can use the layered method to calculate. We divide the triangular area into many layers. As the number of layers increases, the shape of each layer becomes closer to a rectangle. When there are enough layers, each layer can be approximately regarded as a SPRG without thickness, and its equivalent parameters can also be obtained through the previous method, at this time we can find that when $y = -11m, u_x = 0, u_y = a/d * u_0, g = d/a * g_0$, when $y = -6m, u_x = 0, u_y = u_0, g = g_0$, and the parameter change of each layer is continuous, because the distance between the isosceles triangle parts of the circular structure of the two Luoyang bridges changes according to a linear function. Therefore, the water depth in $y$ direction of its equivalent water layer should also change according to a linear function. Therefore, the water depth in $y$ direction of its equivalent water layer should also change according to a linear function. We can then write down an approximate reduced water depth expression $\vec{u}$ and gravitational acceleration expression $g$ of the equivalent anisotropic gradient water layer of the whole Luoyang Bridge. When $-11m < y < -6m$, $u_y = \frac{(d+y+s_1+1)}{d} * u_0$, and because the propagation wave number of the water wave in the $y$ direction in the equivalent gradient anisotropic water layer should be equal to $k_0$, that is, $\sqrt{gu_y} = 1$, so the gravitational acceleration in this area is $g = \frac{d}{(d+y+s_1+1)} * g_0$. And in the $-11m < y < -6m$, as long as the width of isosceles triangle parts of the circular structure is not equal to 0, the flow of water can only propagate along the $y$ direction, and cannot pass through the $x$ direction in the Luoyang Bridge structure. Therefore, the water depth in the $x$ direction is always equal to 0. When $-11m < y < -6m, u_x = 0$. Finally, we can write down the whole approximate reduced water depth expression $\vec{u}$ and gravitational acceleration expression $g$ of the equivalent gradient anisotropic water layer of the Luoyang Bridge as follows:

$$\vec{u}, g = \begin{cases} u_x = 0, u_y = \frac{(d+y+s_1+1)}{d} * u_0, g = \frac{d}{(d+y+s_1+1)} * g_0 & (-11m < y < -6m) \\ u_x = 0, u_y = \frac{a}{d} * u_0, g = \frac{d}{a} * g_0 & (-17m < y < -11m) \end{cases} \quad (S50)$$

This is the equation (10) in the main text.

**Section Two: Experimental Realization**

**(1) Experimental set up**
The experimental set up is shown in Supplementary Figure 6a. The experiment is carried out in a transparent water tank. The water tank is 124cm long and 73cm wide. The point source with angular momentum is generated by the middle motor and propeller. The wavelength of the water wave can be changed by changing the rotation period of the motor. The orange part below the motor is the reduced metagrating model of Luoyang Bridge mentioned in the main text. It is fabricated with a 3D printer. The material is PLA plastic. It is an impermeable rigid body and can completely reflect water waves.

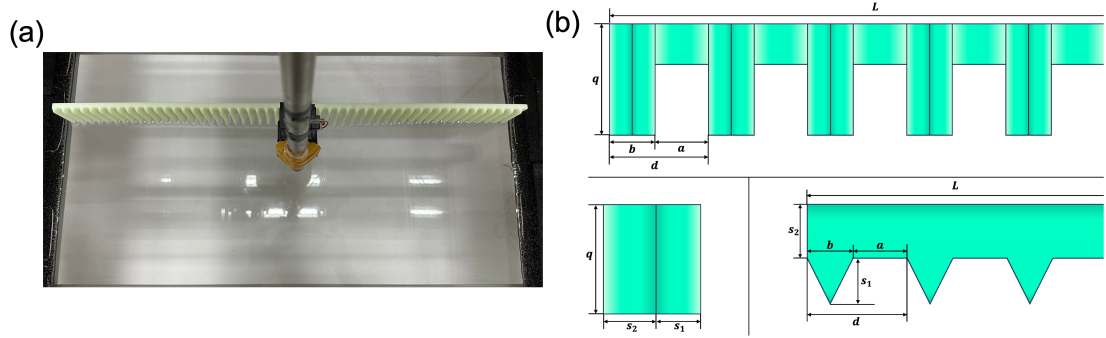

**Supplementary Figure 6.** Experimental equipment structure diagram. (a) Experimental set up diagram. (b) Schematic diagram of the reduced structure of Luoyang Bridge.

The enlarged view of the metagrating is shown in Supplementary Figure 6c, and the size is $L = 64\text{cm}, q = 5cm, d = 1.1cm, a = 0.5cm, b = 0.6\ cm, s_1 = 0.5cm, s_2 = 0.6cm$. Considering the shallow water approximation of water wave equation, the water intake depth $h$ is 7mm in our experiment. The black parts on the left and right sides are the wave elimination device, which can greatly reduce the reflection of water waves of outer boundaries. The specific structure of the wave eliminator is shown in Supplementary Figure 7.

**(2) Wave elimination device**

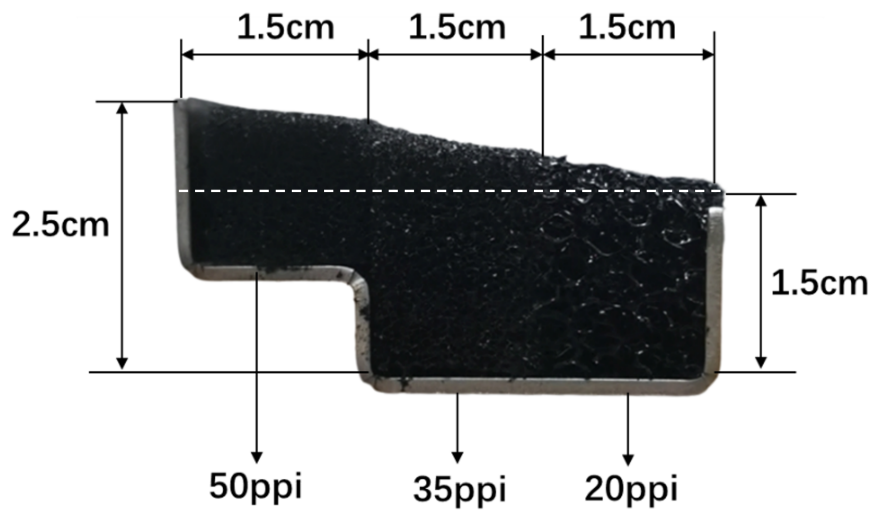

**Supplementary Figure 7.** Cross sectional view of the water elimination device. It is divided into three layers, and the pore size of the sponge increases from left to right [43].

In this experiment, we use the same wave elimination device as before [43], which is an inclined plane composed of three layers of sponge. Both inclined plane and the pores in the sponge can help reduce the energy of the water wave, thereby reducing or even eliminating the reflection of the water wave at the boundary, making the experimental results more accurate. Its specific structure and size are shown in Supplementary Figure 7. The water wave elimination device can be divided into two areas, one is a triangular area with a slope above the white dotted line, the height is 1cm, bottom edge length is 4.5cm, the other is a square area below the white dotted line, the height is 1.5cm, the length is 4.5cm, and the thickness of each layer of sponge is 1.5cm.

The density of pores in each layer of sponge is different, from left to right are 50ppi, 35ppi and 20ppi. Because a small piece of glass is added at the boundary of the water cylinder for our experiment to enhance the air tightness, there is a square notch on the left of Supplementary Figure 7, which can make the wave elimination device fit better with the water cylinder.

**(3) Indicating float ball**

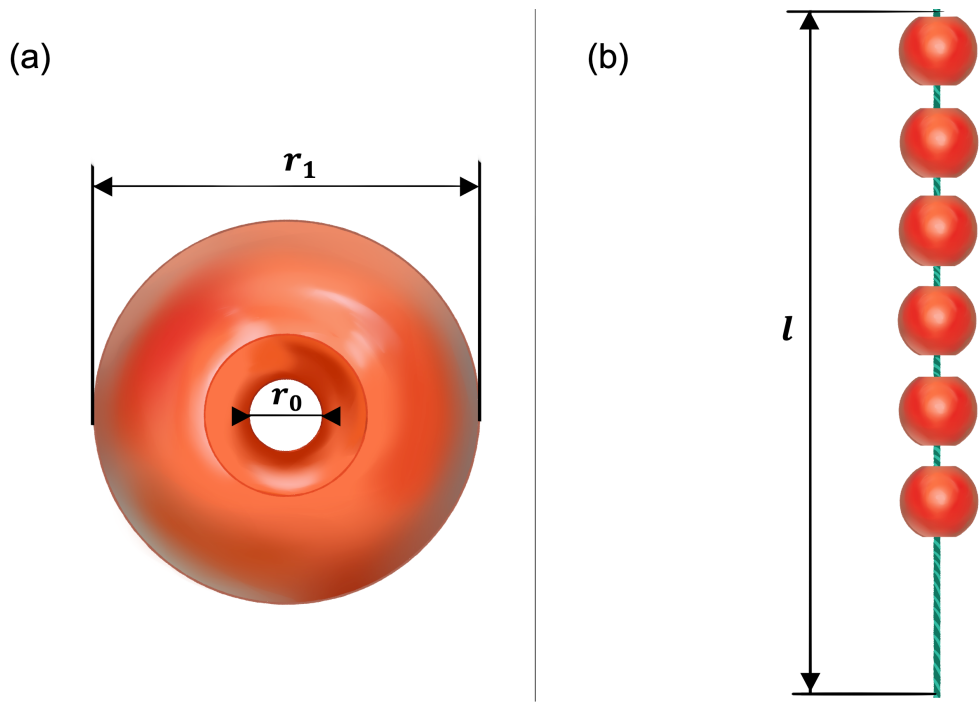

**Supplementary Figure 8.** Schematic diagram of the buoy consisting of float ball. (a) Structure diagram of single float ball. (b) Structure diagram of the buoy consisting of float ball.

The float ball is in orange and is made of ABS plastic. The specific dimensions are shown in Supplementary Figure 8a, the outer diameter of the float ball $r_1$ is 200mm and the inner diameter $r_0$ is 20mm, which can provide 4kg buoyancy at most. Then we connect six float balls in series with nylon rope to form a buoy consisting of float balls. The interval between each float ball is 50cm. The total length of the buoy composed of six float balls and nylon rope $l$ is 7m. As shown in Supplementary Figure 8b, the buoy is hung at the apex of the triangular part of Luoyang Bridge, as for the symmetrical distribution of the wave source center, as shown in Fig. 3d of the main text, the red dot indicates the hanging position of the buoy consisting of float balls, and the red arrow indicates the position of the buoy consisting of float balls.

**(4) The field experiment of Luoyang Bridge**

We also perform experiments near the real Luoyang Bridge, the water depth is about 3m. Here, we use a ship to rotate in situ to excite a vortex water wave. Due to the complex water conditions near Luoyang Bridge, we slow down the frequency of the ship's rotation to 0.09Hz (if the ship continues to slow down, it will be difficult to generate vortex waves), and make the wavelength as large as possible so that it is close to 47m, the experiment near Luoyang Bridge can only be roughly observed. At the same time, in order to distinguish from the results in the laboratory, we changed the rotation direction of the ship to excite a clockwise rotating water wave.

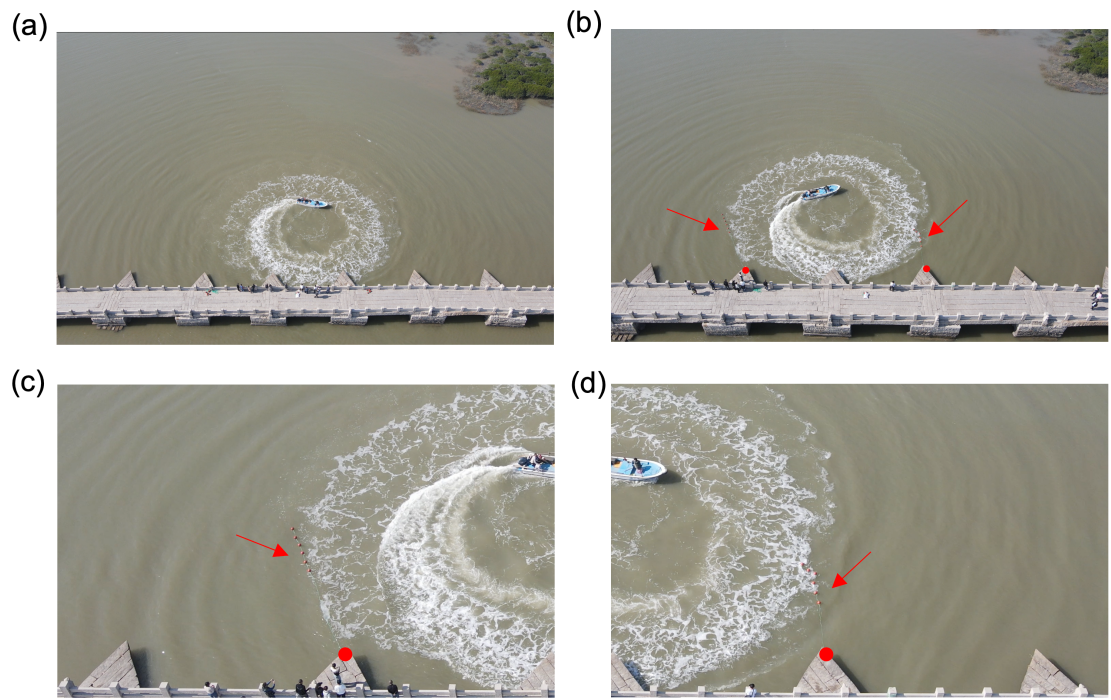

**Supplementary Figure 9.** Field experimental results. (a) Field pattern for the case when the rotating boat is close (6 m) to the real Luoyang Bridge. The boat is clockwise rotating. (b) Field pattern for the case when the rotating boat is close (6 m) to the real Luoyang Bridge with orange float ball. The boat is clockwise rotating. Enlarged image on the left part (c) and right part (d) of Fig. 3d.

The results are shown in Supplementary Figure 9a and b. The wave source center is set 6 m away from Luoyang Bridge, it can be seen that the water wave on the right side of the wave source is weakened to form a unidirectional water surface wave propagating to the left. In order to make the unidirectional water surface wave more obvious, we put two strings of orange float balls on both sides of the wave source. It can be seen that the included angle between the float ball and the vertical direction on the left side is greater than that on the right side. In order to make the deflection angle of the orange float balls look clearer, an enlarged view of the two parts of the orange float balls in Supplementary Figure 9c are shown in Supplementary Figure 9d and f. It can be clearly seen that the left float balls have a larger included angle with the vertical direction, which prove that the strength of the water surface wave propagating to the left is greater than that propagating to the right, i.e. it forms a unidirectional water surface wave propagating to the left.

**(5) Video demonstration of experimental results**

Supplementary Figure 10a (corresponding to Fig. 3e in the main text) shows the clockwise rotating vortex wave excites a unidirectional water surface wave propagating to left near the metagrating. Supplementary Figure 10b (corresponding to Fig. 3f in the main text) shows the counterclockwise rotating vortex wave excites a unidirectional water surface wave propagating to right near the metagrating. Both Supplementary Figure 9a and b have the corresponding video demonstrations, where unidirectional water surface waves can be clearly seen. Supplementary Figure 10c (corresponding to Supplementary Figure 9a) shows clockwise rotating vortex wave excites a unidirectional water surface wave propagating to the left near the real Luoyang Bridge, Supplementary Figure 10d (corresponding to Supplementary Figure 9b) shows that using buoys consisting of float ball to make the unidirectional water surface waves (propagating to left) excited by Luoyang Bridge look more intuitive.

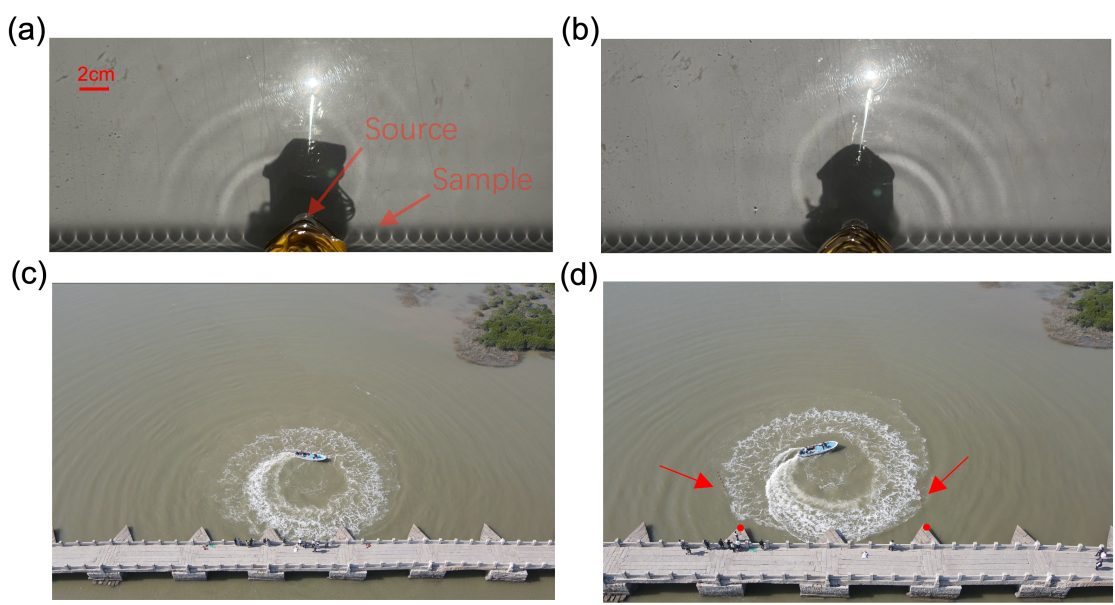

**Supplementary Figure 10.** Demonstration of experimental results. (a) Schematic diagram of video 1, unidirectional water wave propagating to the left excited by the reduced structure of Luoyang Bridge. (b) Schematic diagram of video 2, unidirectional water wave propagating to the right excited by the reduced structure of Luoyang Bridge. (c), Schematic diagram of video 3, unidirectional water wave propagating to the left excited by the real Luoyang Bridge. (d) Schematic diagram of video 4, unidirectional water wave propagating to the left (indicated by red buoys) excited by the real Luoyang Bridge.

## Funding

This work was supported by National Natural Science Foundation of China (Grants Nos. 12374410 and 92050102); National Key Research and Development Program of China (Grants No. 2020YFA0710100); Fundamental Research Funds for the Central Universities (Grants Nos. 20720220033 and 20720230102).

## Reference

*kenyon@xmu.edu.cn